\documentclass[aip,pra,twocolumn,longbibliography]{revtex4}
\usepackage[usenames,dvipsnames]{color}
\usepackage{graphicx}
\usepackage{dcolumn}
\usepackage{bm}
\usepackage{amsmath}
\usepackage{amsfonts}
\usepackage{psfrag}
\usepackage[bookmarks,colorlinks=false]{hyperref}
\usepackage{mathtools}
\usepackage{accents}

\usepackage[mathcal]{euscript}

\begin{document}

\title{Nonlocal Scattering Matrix Description of Anisotropic Polar Heterostructures}

\author{Christopher R. Gubbin}
\author{Simone De Liberato}
\email[Corresponding author: ]{s.de-liberato@soton.ac.uk}
\affiliation{School of Physics and Astronomy, University of Southampton, Southampton, SO17 1BJ, United Kingdom}

\begin{abstract}
Polar dielectrics are a promising platform for mid-infrared nanophotonics, allowing for nanoscale electromagnetic energy confinement in oscillations of the crystal lattice. We recently demonstrated that in nanoscopic polar systems a local description of the optical response fails, leading to erroneous predictions of modal frequencies and electromagnetic field enhancements. In this Paper we extend our previous work providing a scattering matrix theory of the nonlocal optical response of planar, anisotropic, layered polar dielectric heterostructures. The formalism we employ allows for the calculation of both reflection and transmission coefficients, and of the guided mode spectrum. We apply our theory to complex AlN/GaN superlattices, demonstrating the strong nonlocal tuneability of the optical response arising from hybridisation between photon and phonon modes. The numerical code underlying these calculations is provided in an online repository to serve as a tool for the design of phonon-based mid-infrared optoelectronic devices.
\end{abstract}
\maketitle

\section{Introduction}
\label{SecIntro}
Polar dielectric crystals support localised optical modes termed surface phonon polaritons which have recently emerged as a promising platform for mid-infrared nanoscale capture and control of light \cite{Caldwell2015a}. These modes allow for deep sub-wavelength control of light by hybridising it with the optic phonon modes of a polar lattice \cite{Hillenbrand2002}. The resulting modes are morphology dependant, making them highly tuneable \cite{Spann2016,Gubbin2016,Ellis2016,Gubbin2017,Passler2018, Dai2019} with potential applications in nonlinear optics \cite{Gubbin2017b, Gubbin2017c, Razdolski2018}, design of infrared thermal emitters \cite{Greffet2002,Wang2017,Shin2019} and design of nanophotonic circuitry \cite{Li2016,Li2018,Chaudhary2019, Passler2020}.\\
As polar nanophotonics matures, greater interest is being placed on the use of of multi-material systems, exploiting the mature technologies available for fabrication of semiconductor heterostructures. Recent works have demonstrated strong changes in the response of 4H-SiC nanoresonators coated in nanoscopic layers of AlN or Al$_2$O$_3$ \cite{Berte2018}. Crystal hybrid nanostructures in which the entire active region is a superlattice comprised of alternating nanoscopic layers were also highlighted as a highly tuneable platform for mid-infrared nanophotonics \cite{Ratchford2019}.\\
In systems containing truly nanoscopic features local theories have been shown to provide an inaccurate description of the optical response \cite{Ratchford2019, Gubbin2019, Gubbin2020b, Gubbin2020a}. This is because a local model takes lattice phonons to be dispersionless, existing at fixed frequency for all wavevectors. These infinite-mass phonons don't propagate, allowing for a $\delta$-like distribution of screening charges at the crystal surface. In reality optical phonons are dispersive: their frequency changes as a function of wavevector. When considering this spatial dispersion screening charges induced at the crystal surface radiate finite-wavevector longitudinal and transverse optic phonons in each layer. In true nanoscopic layers these phonon modes are quantised and have been shown to lead to additional features in the optical response, frequency shifts, and a decrease in the field confinement \cite{Gubbin2020b, Gubbin2020a}. These effects are analogous to those seen in plasmonic systems \cite{Ciraci2012, Ciraci2013, Mortensen2014}, even though the negative dispersion of optical phonons can lead to a rather different phenomenology.\\

Polar nonlocality is particularly important due to the ease of fabrication of dielectric superlattices comprised of many planar nanoscopic polar layers \cite{Ratchford2019}. In the framework of a local dielectric theory of  the optical response these structures are typically described with a transfer matrix formalism,  considering up- and down-propagating photonic modes in each layer \cite{Berreman1972, Yeh1979, Li1988, Lin-Chung1984, Xu2000, Passler2017, Passler2020smm}. In this model longitudinal and transverse phonons exist at fixed frequencies, and are solely considered through the transverse dielectric tensor of the lattice. When nonlocal effects are taken into account, spatial dispersion means finite-wavevector longitudinal and transverse phonon modes in each layer must also be considered, necessitating application of additional boundary conditions at each interface \cite{Pekar1959,RidleyBook}. We recently developed a general nonlocal theory of polar dielectrics \cite{Gubbin2020b}, which was able to replicate the experimental results obtained in the study of crystal hybrids \cite{Ratchford2019}. Such a theory explained the unexpected observed features as consequences of the nonlocal nature of the optical response. Its isotropic nature made nevertheless impossible to describe the features in the experimental data due to the anisotropic nature of the crystal structure.
\\
This Paper develops our theory of nonlocality in polar planar heterostructures, extending it to account for uniaxial and biaxial crystal anisotropy, important for accurate description of the optical properties of technologically relevant systems. The theory is then exploited to investigate the nonlocal phenomenology and optical tuneability of various superlattices.\\
The rest of this work is structured as follows: in Section \ref{SecTheory}A we develop the theory for anisotropic media, showing the need to expand both the photon and phonon mode space compared to previous studies of isotropic nonlocality \cite{Gubbin2020b, Gubbin2020a}. As in nonlocal systems the transfer matrix formalism is unstable \cite{Passler2017, Passler2020smm}, we employ a scattering matrix approach \cite{Li1996} to calculate the optical response of heterostructures containing arbitrary numbers of layers as discussed in in Section \ref{SecTheory}B. The result is a flexible theory able to calculate the response of complex structures to plane-wave illumination, in addition to finding the spectrum of guided surface phonon polariton modes below the light line. In Section \ref{SecSimulations} we numerically investigate the mid-infrared optical response of GaN/AlN superlattices, showing the emergence of nonlocal branches due to coupling to longitudinal phonon modes in the structure. Furthermore, we consider more complex heterostructures comprised of multiple layer thicknesses which allow fine tuning of the polariton spectrum, something our approach is able to consider with ease. Conclusions and perspectives are drawn in Section \ref{SecConclusions}\\
Finally, our code is provided in an online repository in order to facilitate the exploration of optical nonlocal effects in planar polar heterostructures \cite{Code}.

\section{Nonlocal Theory}
\label{SecTheory}
\subsection{Homogeneous dielectrics}
The optical response of a planar heterostructure is found by relating the amplitude of an external illumination to those of fields reflected and transmitted by the heterostructure. This entails applying electromagnetic and mechanical boundary conditions at each interface in the planar system to uniquely determine the mode amplitudes in each of the structure's layers. 
\begin{figure}
\includegraphics[width=0.3\textwidth]{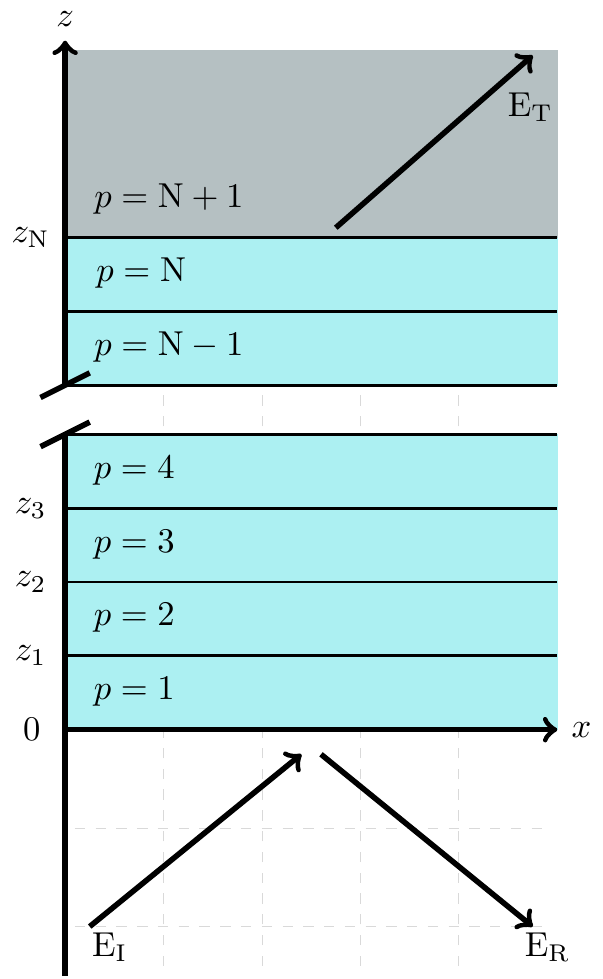}
\caption{\label{fig:Fig0} Sketch of the system under study. This consists of a planar stack of layers numbered from $p=1$ to $N$ on a deep substrate labelled $p=N + 1$. The growth axis of the heterostructure is along $z$ and the $p=1$ layer begins at $z=0$. The heterostructure is illuminated from below in the $xz$ plane and is homogeneous along $y$.}	
\end{figure}\\
We consider a coordinate system where the $xy$ plane is aligned with the heterostructure interfaces and $z$ is parallel to the growth-axis. We let $z=0$ at the interface between the first layer of the heterostructure and the incident medium. Layers of the heterostructure are numbered $p = 1$ to $p = N$, while the incident medium and substrate are numbered $p = 0$ and $p = N + 1$ respectively as illustrated in Fig.~\ref{fig:Fig0}. The thickness of each layer $p$ is $d_p$. Without loss of generality we consider fields propagating in the $xz$ plane. The wavevector in layer $p$ is thus
\begin{equation}
	\mathbf{k}_p = k_0 \left(\zeta, 0, q_{p}\right), \label{eq:Wavevector}
\end{equation}
where $k_0$ is the free-space wavevector, the dimensionless in-plane wavevector $\zeta$ is conserved between layers and the dimensionless out-of-plane wavevector is $q_p$. \\
Polar heterostructures are normally fabricated by epitaxy, so one of the crystal axis aligns with the growth ($z$) direction. We can thus restrict ourselves to media whose properties are diagonal in the lab-frame, thus allowing us to treat crystal structures with both uniaxial and biaxial anisotropy. Under these conditions in the local case crystals whose symmetry axis align with the cartesian frame, support two possibly degenerate photonic modes: the ordinary and extraordinary rays \cite{Passler2017}.\\
Nonlocality is accounted for in our model through effective stress tensor $\bar{\tau}$, defined in relation to the lattice strain
\begin{equation}
	s_{k,l} = \frac{1}{2} \left[ \frac{\partial X_k}{\partial r_l} + \frac{\partial X_l}{\partial r_k} \right], \label{eq:strain}
\end{equation}
where $r_{l=\{1,2,3\}}$ is the spatial coordinate $\{x,y,z\}$ and $X_l$ is a component of the relative ionic displacement in the crystal. The tensor $\bar{\tau}$ is given by
\begin{equation}
	\tau_{i,j} = c_{i,j,k,l} s_{k,l}, \label{eq:elacons}
\end{equation}
where $c_{i,j,k,l}$ is the fourth-rank elasticity tensor of the lattice. The non-zero components of $c_{i,j,k,l}$ are derived considering the point group of the lattice \cite{Bona2004}, and are given in Appendix A for cubic and hexagonal lattices. \\
The key ingredient in a nonlocal model is spatial dispersion, or how the longitudinal and transverse phonon frequencies of the crystal lattice vary as functions of the excitation wavevector. Although the exact phonon dispersion can be a priori calculated with {\it ab inito} simulations or measured by inelastic scattering, its functional form is often too complex for practical use. In order to provide a semi-analytic and practical theory, we relied instead on a quadratic approximation which we demonstrated can provide accurate results even for nanometer-sized structures \cite{Gubbin2020b}. 
In our model phonon dispersions are thus of the form
\begin{equation}
	\omega_i^2 = \omega_{i, 0}^2 + \left(\boldsymbol{\beta}_i\cdot \mathbf{k}\right)^2, \label{eq:quaddisp}
\end{equation}
where $\omega_{i, 0}$ is the zero-wavevector frequency of phonon band $i$ and $\boldsymbol{\beta}_i$ is a polarisation dependent vector of nonlocal velocities \cite{Gubbin2020b}. Note that in an anisotropic lattice $i$ is a composite index, encompassing both polarisation state and phonon branch. \\
The effect of nonlocality can be understood considering the dispersion of the ordinary ray, polarised along the $y$-direction by Eq.~\ref{eq:Wavevector}. Locally this obeys the standard dispersion relation
\begin{equation}
	q_p^2 = \epsilon_{p}^{\text{L},y}\left(k_0\right) - \zeta^2
\end{equation}
 where $\epsilon_{p}^{\text{L},y}$ is the $y$ component of the layer's local dielectric function. This equation yields a single up- and down-propagating out-of-plane wavevector $q_p$ for each frequency $\omega = c k_0$, which correspond to propagative or evanescent excitations depending on the sign of $\epsilon_{p}^{\text{L},y}$ and magnitude of $\zeta$. Accounting for nonlocality the phonon frequencies entering the dielectric function are wavevector dependent through Eq.~\ref{eq:quaddisp}, yielding a dispersion relation of the form 
\begin{equation}
	q_p^2 = \epsilon_{p}^{\text{NL},y} \left(k_0, q_p, \zeta \right) - \zeta^2,
\end{equation}
where $\epsilon_{p}^{\text{NL},y}$ is now the $y$ component of the wavevector-dependent nonlocal dielectric function. For each frequency there are now two unique up- and down-propagating solutions which can be interpreted as an ordinary photon and ordinary TO phonon. The same is true for the extraordinary ray, whose electric field is polarised in the $xy$ plane. The crystal also supports a single longitudinal phonon mode. This means each layer supports 5 distinct up- and down-propagating modes, {\it a priori} all non-degenerate for an anisotropic crystal, and a generic wave propagating through the superlattice will thus be determined by 10 independent modal amplitudes.

To find the out-of-plane wavevectors and amplitudes of the 5 modes in the homogeneous lattice we utilise the equations of motion of the lattice coupled to the electromagnetic fields \cite{Gubbin2019}. The electric and magnetic fields $\left(\mathbf{E} \; \& \; \mathbf{H}\right)$ obey Ampere's law and the Maxwell-Faraday equation
\begin{align}
	\nabla \times \mathbf{H} &= - i \omega\left[\epsilon_0 \mathbf{E} + \mathbf{P}\right], \label{eq:Amp} \\
	\nabla \times \mathbf{E} &=  i \omega \mu_0 \bar{\mu} \mathbf{H},
\end{align}
where $\bar{\mu}$ is the permeability tensor, we assumed $e^{-i \omega t}$ time dependance, and $\mathbf{P}$ is the macroscopic polarisation. Note that we use bold symbols for vectors and barred ones for matrices. The electric and polarisation fields relate to the relative ionic displacement $\mathbf{X}$ through constitutive relation
\begin{equation}
	\mathbf{P} = \bar{\alpha} \mathbf{X} + \epsilon_0\left(\bar{\epsilon}_{\infty} - \bar{\mathrm{I}}\right) \mathbf{E}, \label{eq:cons}
\end{equation} 
where $\bar{\epsilon}_{\infty}$ is the high-frequency permittivity tensor, $\bar{\alpha}$ is the effective charge density of the lattice and $\bar{\mathrm{I}}$ is the identity matrix. Finally we consider how the electric field couples to the polar lattice, described in the continuum limit by matrix equation 
\begin{equation}
	\left[\bar{\omega}_{\mathrm{T}}^2  - \omega\left(\omega + i \bar{\gamma}\right) \right] \mathbf{X} + \nabla \cdot \bar{\tau} =\frac{\bar{\alpha}}{\rho} \mathbf{E}, \label{eq:IonEOM}
\end{equation}
where $\mathbf{X}$ is the relative ionic displacement, $\bar{\omega}_{\mathrm{T}}$ describes the TO phonon frequencies along the different axis, $\bar{\gamma}$ their damping rates, and $\rho$ is the effective mass density. The quantity $\bar{\tau}$ is the effective stress tensor for optical phonons, which can be derived for arbitrary lattice symmetry, considering non-zero components of the lattice stress tensor as explained in Appendix A \cite{Trallero-Giner1992}, where explicit forms are also provided for the most technologically relevant cubic and hexagonal cases.\\

Although such explicit forms are provided for completeness, and can be straightforwardly (albeit cumbersomely) implemented if needed, we can notice that for the planar systems object of this work the isotropic version of the stress tensor can be safely used also for anisotropic crystals. The in-plane wavevector $k_0 \zeta$ is in fact determined by the photonic field and conserved across all layers of the stack. Its magnitude  is thus negligible compared to the lattice wavevector, meaning that only the unconstrained wavevector component in the $z-$direction $k_0 q$ will affect the nonlocal response. There are no nonlocal effects in the plane and the only relevant velocities are those parallel to the growth direction. The isotropic form for $c_{ijkl}$, parametrised by the longitudinal ($\beta_{\mathrm{L}}$) and transverse ($\beta_{\mathrm{T}}$) velocities in the $z-$direction thus provides quantitatively correct results, while also leading to a much simpler algebra \cite{Trallero-Giner1998a}
\begin{widetext}
\small
\begin{equation}
\label{eq:Tau}
  \bar{\tau} = \left( \begin{array}{ccc}
  \beta_{\mathrm{L}}^2 \partial_x \mathrm{X}_x + \left(\beta_{\mathrm{L}}^2 - 2 \beta_{\mathrm{T}}^2 \right) \left( \partial_y \mathrm{X}_y + \partial_z \mathrm{X}_z\right)  & \beta_{\mathrm{T}}^2 \left(\partial_y \mathrm{X}_x + \partial_x \mathrm{X}_y\right) & \beta_{\mathrm{T}}^2 \left(\partial_z \mathrm{X}_x + \partial_x \mathrm{X}_z\right)  \\
  \beta_{\mathrm{T}}^2 \left(\partial_x \mathrm{X}_y + \partial_y \mathrm{X}_x\right) & \beta_{\mathrm{L}}^2 \partial_y \mathrm{X}_y + \left(\beta_{\mathrm{L}}^2 - 2 \beta_{\mathrm{T}}^2 \right) \left( \partial_x \mathrm{X}_x + \partial_z \mathrm{X}_z\right) & \beta_{\mathrm{T}}^2 \left(\partial_z \mathrm{X}_y + \partial_y \mathrm{X}_z\right)\\
  \beta_{\mathrm{T}}^2 \left(\partial_x \mathrm{X}_z + \partial_z \mathrm{X}_x\right) & \beta_{\mathrm{T}}^2 \left(\partial_y \mathrm{X}_z + \partial_z \mathrm{X}_y\right) & \beta_{\mathrm{L}}^2 \partial_z \mathrm{X}_z + \left(\beta_{\mathrm{L}}^2 - 2 \beta_{\mathrm{T}}^2 \right) \left( \partial_x \mathrm{X}_x + \partial_y \mathrm{X}_y\right)
\end{array} \right).
\end{equation}
\normalsize
\end{widetext}
The ionic equation of motion in Eq.~\ref{eq:IonEOM} and the nonlocal stress tensor in Eq.~\ref{eq:Tau} yield phonon dispersions in the form of Eq.~\ref{eq:quaddisp}. For example for the ordinary phonon obeys Eq.~\ref{eq:Wavevector} with $\mathrm{X}_x = \mathrm{X}_z = 0$, leading to
\begin{equation}
	\nabla \cdot \bar{\tau} =  - \beta_\mathrm{T}^2 k^2 \left( \begin{array}{c}
  0\\
  1\\
  0
\end{array} \right) \mathrm{X}_y.
\end{equation}
The ordinary phonon dispersion is then deduced from inspection of Eq.~\ref{eq:IonEOM} 
\begin{equation}
	\omega_{\mathrm{T}, y}^2 = \omega_{\mathrm{T}, y, 0}^2 - \beta_{\mathrm{T}}^2 k^2,
\end{equation}
meaning that in Eq.~\ref{eq:quaddisp} for this polarisation
\begin{equation}
	\boldsymbol{\beta}_i = \beta_{\mathrm{T}} \left( \begin{array}{c}
  1\\
  0\\
  1
\end{array} \right).
\end{equation}\\
The system in Eqs.~\ref{eq:Amp}-\ref{eq:IonEOM} can be written in the form of a $12 \times 12$ matrix equation 
\begin{equation}
	\bar{\mathrm{M}} \mathbf{F} = 0, \label{eq:Full}
\end{equation}
where the coefficient vector is $ \mathbf{F} = \left[\mathbf{E, H, P, X}\right]$ and the matrix $\bar{\mathrm{M}}$ can be written in block form as
\begin{equation}
	\bar{\mathrm{M}} = \left(\begin{array}{cccc}
	c \epsilon_0 \bar{\mathrm{I}} & \bar{\psi} &  c \bar{\mathrm{I}} & \bar{0}\\
	-\bar{\psi} & c \mu_0 \bar{\mu} & \bar{0} & \bar{0}\\
	\epsilon_0\left(\bar{\epsilon}_{\infty} - 1\right) & \bar{0} & - \bar{\mathrm{I}} & \bar{\alpha} \\
	\frac{\bar{\mu}}{\rho} & \bar{0} & \bar{0} & \omega \left(\omega + i \bar{\gamma}\right) - \bar{\omega}_{\mathrm{T}}^2 - \bar{\eta}
\end{array}
 \right), \label{eq:fullmatrix}
\end{equation}
where $c$ is the speed of light, $\bar{\mathrm{I}}$ is a $3 \times 3$ identity matrix, we defined $\bar{\eta} = \nabla \cdot \bar{\tau}$, and the curl operator in Fourier space can be written with the parameters defined in Eq.~\ref{eq:Wavevector}
\begin{equation}
	\bar{\psi} =  \left( \begin{array}{ccc}
 	0 & - q & 0 \\
 	q & 0 & - \zeta \\
 	0 & \zeta & 0
 \end{array}\right).
\end{equation}
The explicit form of the non-zero components of $\bar{\eta}$ can be found in Appendix B.\\
The $12 \times 12$ system in Eq.~\ref{eq:Full} would generally have $12$ unique solutions. As described in Section \ref{SecTheory}A though, we restrict ourselves to the relevant physical case of diagonal material tensors ($\bar{\alpha},\bar{\epsilon}_{\infty},\bar{\gamma},\bar{\omega}_T$), leading to only 5 distinct solutions. This means Eq.~\ref{eq:Full} contains redundancies. To proceed we eliminate $\mathrm{E}_z, \mathbf{H}, \mathbf{P}$ yielding a $5 \times 5$ equation set as detailed in Appendix C. The resulting equation is quadratic in $q$ and can be written
\begin{equation}
	\left[q^2 \bar{\mathrm{A}} + q \bar{\mathrm{B}} + \bar{\mathrm{C}} \right] \mathbf{F}' = 0, \label{eq:redeom}
\end{equation}
where the reduced coefficient vector is $\mathbf{F}' = \left[\mathrm{E}_x, \mathrm{E}_y, \mathbf{X}\right]$ and $\bar{\mathrm{A}}, \bar{\mathrm{B}},\bar{\mathrm{C}}$ are $5 \times 5$ matrices given in full in Appendix C. To arrive at a linear algebraic equation for $q$, suitable for implementation in a numerical eigensolver, we use auxiliary vector $q \mathbf{F}'$ to write
\begin{align}
	q \left(\begin{array}{cc}
	\bar{\mathrm{B}} & \bar{\mathrm{A}} \\
	\bar{\mathrm{I}} & 0 	
\end{array}\right) \left(\begin{array}{c} \mathbf{F}' \\ q \mathbf{F}' \end{array}\right) + \left(\begin{array}{cc}
	\bar{\mathrm{C}} & 0 \\
	0 & - \bar{\mathrm{I}}	
\end{array}\right)\left(\begin{array}{c} \mathbf{F}' \\ q \mathbf{F}' \end{array}\right) = 0. \label{eq:lineom}
\end{align}
Here the lower block of the matrix equation enforces the functional relation between the eigenvalue $q$ and $\mathbf{F}'$ on the top and $q \mathbf{F}'$ on the bottom row of the solution vector. Now we define $\boldsymbol{\Psi} = \left[ \mathbf{F}', q \mathbf{F}'\right]^{\mathrm{T}}$ to arrive at the linear eigenequation
\begin{equation}
	q \boldsymbol{\Psi} = - \bar{\Upsilon} \boldsymbol{\Psi}, \label{eq:eig}
\end{equation} 
where $\bar{{\Upsilon}}$ is a $10 \times 10$ matrix whose non-zero elements are given in Appendix C. Solving Eq.~\ref{eq:eig} yields the out-of-plane wavevector in the nonlocal dielectric and the reduced coefficient vector $\mathbf{F}'$, from which the full field vector $\mathbf{F}$ can be reconstructed utilising the relations also provided in Appendix C.\\
The eigenvalue problem in Eq.~\ref{eq:eig} yields 10 solutions of which 5  are up-propagating and 5 down-propagating. These are sorted according to mode type and polarisation state as described in Appendix D, yielding vector $\left[\mathbf{u, d}\right]$ where the vector of up-propagating modes is $\mathbf{u} = \left[u_{\mathrm{ph}}^{\mathrm{TE}}, u_{\mathrm{ph}}^{\mathrm{TM}},u_{\mathrm{T}}^{\mathrm{TE}}, u_{\mathrm{T}}^{\mathrm{TM}}, u_{\mathrm{L}}\right]$ and that for down-propagating modes is $\mathbf{d} = \left[d_{\mathrm{ph}}^{\mathrm{TE}}, d_{\mathrm{ph}}^{\mathrm{TM}},d_{\mathrm{T}}^{\mathrm{TE}}, d_{\mathrm{T}}^{\mathrm{TM}}, d_{\mathrm{L}}\right]$, where TM and TE superscripts describe the mode polarization and the the ph, T, and L subscripts stand for photon, TO, and LO phonon respectively.

\subsection{Planar Heterostructures}
\begin{figure}
\includegraphics[width=0.4\textwidth]{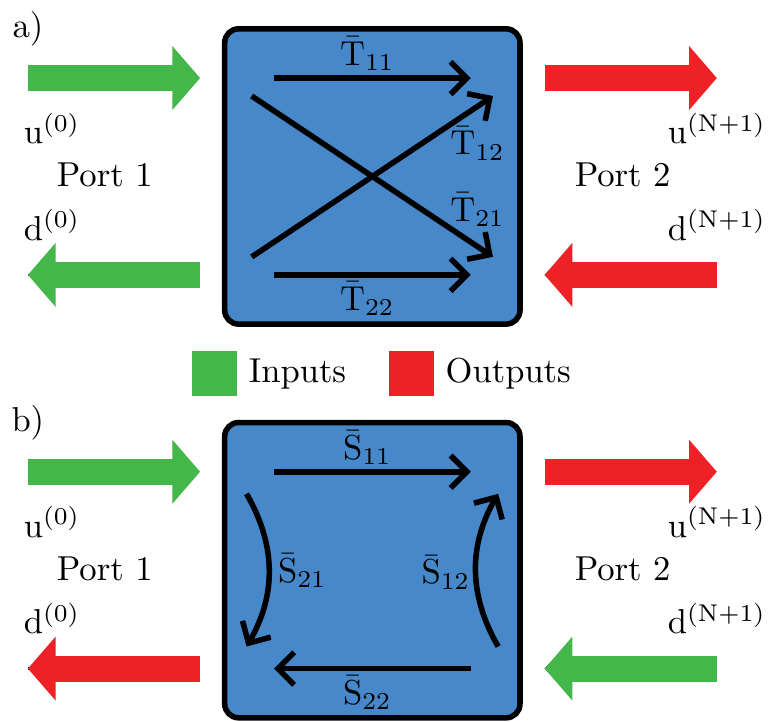}	
\caption{\label{fig:Fig00} A comparison of transfer matrix and scattering matrix algorithms. Inputs (items to be multiplied by the matrix) are indicated by green arrows while outputs are indicated by red arrows. a. Shows a transfer matrix algorithm. The four transfer matrix blocks $\bar{\mathrm{T}}_{i j}$ comprise matrix $\bar{\mathrm{T}}$ in Eq.~\ref{eq:tmmeq}. The algorithm linearly calculates incoming and outgoing fields at Port 2 from the amplitudes of both incoming and outgoing fields at Port 1. b. Shows a scattering matrix algorithm. The four scattering matrix blocks $\bar{\mathrm{S}}_{i j}$ comprise matrix $\bar{\mathrm{S}}$ in Eq.~\ref{eq:smmeq}. The algorithm linearly calculates outgoing fields at both ports from the incoming fields at both ports.}
\end{figure}

\begin{table*}
\def\arraystretch{1.5}
\begin{tabular}{l|ccccccccccc}
  & $\omega_{\mathrm{L}, \parallel}$ & $\omega_{\mathrm{L}, \perp}$ & $\omega_{\mathrm{T}, \parallel}$ & $\omega_{\mathrm{T}, \perp}$ & $\gamma_{\parallel}$ & $\gamma_{\perp}$ & $\epsilon_{\infty, \parallel}$ & $\epsilon_{\infty, \perp}$  & $\beta_{\mathrm{L}}$ & $\beta_{\mathrm{T}}$ \\
  \hline
SiC & $967.7 \mathrm{cm}^{-1}$ & $972.7 \mathrm{cm}^{-1}$ & $783.6 \mathrm{cm}^{-1}$ & $796.6 \mathrm{cm}^{-1}$  & $2 \mathrm{cm}^{-1}$ & $2 \mathrm{cm}^{-1}$ & $6.78$ & $6.56$ & $15.4 \times 10^3 \mathrm{m s}^{-1}$ & $9.2 \times 10^3 \mathrm{m s}^-1$ \\
 AlN & $891.0 \mathrm{cm}^{-1}$ & $912.0 \mathrm{cm}^{-1}$ & $610.0 \mathrm{cm}^{-1}$ & $669.0 \mathrm{cm}^{-1}$ & $6 \mathrm{cm}^{-1}$ & $6 \mathrm{cm}^{-1}$ & $9.28$ & $7.73$ & $5.1\times 10^3 \mathrm{m s}^{-1}$ & $3.0 \times 10^3 \mathrm{m s}^{-1}$ \\
 GaN & $732.5 \mathrm{cm}^{-1}$ & $742.1 \mathrm{cm}^{-1}$ & $537.0 \mathrm{cm}^{-1}$ & $560.0 \mathrm{cm}^{-1}$ & $4 \mathrm{cm}^{-1}$ & $4 \mathrm{cm}^{-1}$ & $5.47$ & $5.42$ & $6.5 \times 10^3 \mathrm{m s}^{-1}$ & $2.9 \times 10^3  \mathrm{m s}^{-1}$ 
\end{tabular}
\caption{\label{fig:Tab1}The parameters used in the simulations. Local parameters for SiC (4H polytype) are taken from \citep{Mutschke1998}. Nonlocal parameters are approximated from the low-wavevector phonon dispersion \cite{Karch1994}. Local paramaters for AlN are taken from \citep{Haboeck2003}. Nonlocal parameters are taken from a previous study \cite{Gubbin2020b}. Additionally the loss is increased in the AlN to better replicate damping through the superlattice layers, as found to be necessary in a prior work \cite{Gubbin2020b}. For GaN local parameters are taken from \citep{Kasic2000} while nonlocal parameters are estimated from the phonon dispersions \citep{Bungaro2000}.}
\end{table*}

We previously considered the eigenmodes of a homogeneous nonlocal dielectric. In order to calculate the optical response of a multilayer stack it is necessary to understand how these modes propagate through the heterostructure, particularly across interfaces between piecewise homogeneous layers. At interfaces between nonlocal layers it is necessary to apply additional boundary conditions, compared to the local case, to fully describe all field components. It has been previously demonstrated in Refs.~\onlinecite{Gubbin2020b, Gubbin2020a} by consideration of the continuity of energy flux across nonlocal interfaces that the appropriate additional boundary conditions are continuity of the relative ionic displacement $\mathbf{X}$ and of the normal component of $\bar{\tau}$ (Eq.~\ref{eq:Tau})
\begin{equation}
	\bar{\tau} \cdot \hat{\mathbf{z}} = \left(\begin{array}{c}
	\beta_\mathrm{T}^2 \left(\partial_z \mathrm{X}_x + \partial_x \mathrm{X}_z\right)	\\
	\beta_\mathrm{T}^2 \left(\partial_z \mathrm{X}_y + \partial_y \mathrm{X}_z\right)\\
	\beta_\mathrm{L}^2  \partial_z \mathrm{X}_z + \left(\beta_{\mathrm{L}}^2 - 2 \beta_{\mathrm{T}}^2 \right) \left( \partial_x \mathrm{X}_x+ \partial_y \mathrm{X}_y\right)
\end{array}\right), \label{eq:normtau}
\end{equation}
whose $xy \; (z)$ components are an effective shear (normal) stress. There are two algorithmic approaches typically employed to calculate propagating fields through a multilayer stack, the transfer matrix approach and scattering matrix approach. These are schematised and compared in Fig.~\ref{fig:Fig00}. In the following we discuss both methods and explain why a scattering matrix approach is required for the nonlocal systems under consideration.\\
\subsubsection{Transfer Matrices}
We can relate the fields in two adjacent piecewise homogeneous layers $p$ and $p+1$, whose intermediate interface is located at $z = z_p$, through
\begin{equation}
	\left[ \begin{array}{c}
 	\mathbf{u}^{(p+1)} \left(z_p^+\right)\\
 	\mathbf{d}^{(p+1)} \left(z_p^+\right)	
 \end{array}\right] = \bar{t}^{(p)} \left[ \begin{array}{c}
 	\mathbf{u}^{(p)} \left(z_p^-\right)\\
 	\mathbf{d}^{(p)} \left(z_p^-\right)	
 \end{array}\right],\label{eq:tmm1}
\end{equation}
where the interface matrix $\bar{t}^{(p)}$, defined in Appendix E, encodes the full set of boundary conditions to be applied in the nonlocal case, that is the continuity of $\mathbf{H}_{\parallel}, \mathbf{E}_{\parallel}, \dot{\mathbf{X}}, \bar{\tau}_{\perp}$. We can write the propagation matrix for a single layer, which links mode amplitudes at the interior of each of its interfaces as
\begin{equation}
	\left[ \begin{array}{c}
 	\mathbf{u}^{(p)}\left(z_p^-\right)\\
 	\mathbf{d}^{(p)} \left(z_p^-\right)	
 \end{array}\right] = \bar{\phi}^{(p)} \left[ \begin{array}{c}
 	\mathbf{u}^{(p)} \left(z_{p-1}^+\right)\\
 	\mathbf{d}^{(p)} \left(z_{p-1}^+\right)	
 \end{array}\right],
\end{equation}
where the propagation matrix in the homogeneous layer is given by
\begin{equation}
	\bar{\phi}^{(p)} = \left( \begin{array}{cc}
 	e^{i \bar{q}_z^{(p) u} d_p} & \bar{0}\\
 	\bar{0} & 	e^{i \bar{q}_z^{(p) d} d_p}
 \end{array}\right), \label{eq:propmat}
\end{equation}
where $\bar{q}_z^{(p) u} ( \bar{q}_z^{(p) d})$ is the vector of up-propagating (down-propagating)  out-of-plane wavevectors in layer $p$. Taken together we define a new matrix $\tilde{t}^{(p)} = \bar{t}^{(p)} \bar{\phi}^{(p)}$ and can write 
\begin{equation}
	\left[ \begin{array}{c}
 	\mathbf{u}^{(p+1)} \left(z_p^+\right)\\
 	\mathbf{d}^{(p+1)} \left(z_p^+\right)	
 \end{array}\right] = \tilde{t}^{(p)} \left[ \begin{array}{c}
 	\mathbf{u}^{(p)} \left(z_{p-1}^+\right)\\
 	\mathbf{d}^{(p)} \left(z_{p-1}^+\right)	
 \end{array}\right].
\end{equation}
A transfer matrix algorithm entails repeated applications of this equation through the stack, yielding the output fields
\begin{equation}
	\left[ \begin{array}{c}
 	\mathbf{u}^{(N+1)}\\
 	\mathbf{d}^{(N+1)} 
 \end{array}\right] = \bar{\mathrm{T}}^{(N)} \left[ \begin{array}{c}
 	\mathbf{u}^{(0)}\\
 	\mathbf{d}^{(0)} 	
 \end{array}\right], \label{eq:tmmeq}
\end{equation}
where 
\begin{equation}
	\bar{\mathrm{T}}^{(N)} = \prod_{0 < p \le N} \tilde{t}^{(p)}.
\end{equation}
An illustration of this procedure can be found in Fig.~\ref{fig:Fig00}a, with the input fields in layer $0$ yield outputs in layer $N+1$. Unfortunately transfer matrix algorithms are numerically unstable in the limit of thick absorbing layers as the propagation matrix Eq.~\ref{eq:propmat}, describing how fields at one side of a layer relate to those on the other diverges due to the imaginary part of the modal wavevector \cite{Li1996}. In nonlocal phononics optical wavelengths are typically of order $10\mu$m, while phonon wavelengths are sub-nanometer. This means phonon out-of-plane wavevectors are extremely large, causing divergence even in films as thin as $10$nm. The instability arises from the application of Eq.~\ref{eq:propmat} which results in dramatic exponential growth of the fields and can be overcome by recasting the problem into a scattering matrix formalism.\\
\subsubsection{Scattering Matrices}
In a scattering matrix approach we seek the stack matrix $\bar{\mathrm{S}}^{(N)}$ which relates the incoming and outgoing fields from the heterostructure
\begin{equation}
	 \left[ \begin{array}{c}
 	\mathbf{u}^{(N+1)}\\
 	\mathbf{d}^{(0)} 
 \end{array}\right] = \bar{\mathrm{S}}^{(N)} \left[ \begin{array}{c}
 	\mathbf{u}^{(0)} \\
 	\mathbf{d}^{(N+1)} 
 \end{array}\right]. \label{eq:smmeq}
\end{equation}
The known quantities $\mathbf{u}^{(0)}$ describing the up-propagating modes in the lowest (0th) layer and $\mathbf{d}^{(N+1)}$ describing the down-propagating modes in the highest ($N+1$) layer are on the right-hand side, while the scattered fields, propagating away from the heterostructure, are on the left-hand side.  This is illustrated in Fig.~\ref{fig:Fig00}b. It is useful to recast $\bar{\mathrm{S}}^{(N)}$ in block form 
\begin{equation}
	 \left[ \begin{array}{c}
 	\mathbf{u}^{(N+1)}\\
 	\mathbf{d}^{(0)} 
 \end{array}\right] = \left[\begin{array}{cc}
	\bar{\mathrm{T}}_{u,u}^{(N)} & \bar{\mathrm{R}}_{u,d}^{(N)} \\
	\bar{\mathrm{R}}_{d,u}^{(N)} & \bar{\mathrm{T}}_{d,d}^{(N)}
\end{array}\right]
 \left[ \begin{array}{c}
 	\mathbf{u}^{(0)} \\
 	\mathbf{d}^{(N+1)} 
 \end{array}\right], \label{eq:smm}
\end{equation}
where the notation using $\mathrm{R}, \mathrm{T}$ and the subscripts take the natural meaning of reflection and transmission matrices, for example the transmission matrix that yields the up-propagating waves in layer $p+1$ is $\bar{\mathrm{T}}_{u,u}^{(p)}$. As in a transfer matrix approach we define an interface matrix, $\bar{s}^{(p)}$, linking the fields on each side of the interface between piecewise homogeneous layers
\begin{align}
	\left[ \begin{array}{c}
 	\mathbf{u}^{(p+1)} \left(z_p^+\right)\\
 	\mathbf{d}^{(p)} \left(z_p^-\right)	
 \end{array}\right] &= \bar{s}^{(p)} \left[ \begin{array}{c}
 	\mathbf{u}^{(p)} \left(z_p^-\right)\\
 	\mathbf{d}^{(p+1)} \left(z_p^+\right)	
 \end{array}\right]\nonumber \\
 &=  \left[\begin{array}{cc}
	\bar{t}_{u,u}^{(p)} & \bar{r}_{u,d}^{(p)} \\
	\bar{r}_{d,u}^{(p)} & \bar{t}_{d,d}^{(p)}
\end{array}\right] \left[ \begin{array}{c}
 	\mathbf{u}^{(p)} \left(z_p^-\right)\\
 	\mathbf{d}^{(p+1)} \left(z_p^+\right)	
 \end{array}\right], \label{eq:smmbase}
\end{align}
where as for the full stack matrix $\bar{\mathrm{S}}^{N}$ we recast in block form. The layer scattering matrix $\tilde{s}^{(p)}$ is related to the interface one $\bar{s}^{(p)}$ by
\begin{align}
	\tilde{s}^{(p)} = \left[\begin{array}{cc}
	\bar{\mathrm{I}} & 0 \\
	0 &e^{- i \bar{q}_z^{(p) d} d_p}
\end{array}\right]\bar{s}^{(p)} \left[\begin{array}{cc}
	e^{i \bar{q}_z^{(p) u} d_p} & 0 \\
	0 & \bar{\mathrm{I}}
\end{array}\right], \label{eq:1sm}
\end{align}
in which, due to the change in sign of the wavevector exponent compared to Eq.~\ref{eq:propmat}, the instability of the transfer matrix approach is removed. Furthermore by linear algebra the interface scattering matrix $\bar{s}^{(p)}$ can be related to the blocks of the interface transfer matrix $\bar{t}^{(p)}$ by
\begin{align}
	\bar{s}^{(p)} =  \left[\begin{array}{cc}
	\bar{t}_{1,1}^{(p)} - \bar{t}_{1,2}^{(p)} \bar{t}_{2,2}^{(p) -1} \bar{t}_{2,1}^{(p)} & \bar{t}_{1,2}^{(p)} \bar{t}_{2,2}^{(p) -1} \\
	-\bar{t}_{2,2}^{(p) -1}\bar{t}_{2,1}^{(p)} & \bar{t}_{2,2}^{(p) -1}
\end{array}\right].
\end{align}\\
This approach is more convoluted and less efficient than a transfer matrix method, because in a transfer matrix method the stack matrix $\bar{\mathrm{T}}$ can be computed directly from matrix multiplication of the individual layer matrices $\tilde{t}$, while the stack matrix $\bar{\mathrm{S}}$ must be constructed iteratively. As each new layer is considered equations of form Eq.~\ref{eq:smmbase} are written for the two adjacent layers, combined to eliminate common mode coefficients, and finally rearranged into the form Eq.~\ref{eq:smm} \cite{Li1996, Rumpf2011}. Such a procedure is neatly summarised by the Redheffer star product, which describes the multiplication of scattering matrices \cite{Tervo2001} to find
\begin{align}
	\bar{\mathrm{T}}_{u,u}^{(p)} &= \tilde{t}_{u,u}^{(p)} \left[ \bar{\mathrm{I}} - \bar{\mathrm{R}}_{u,d}^{(p - 1)} \tilde{r}_{d,u}^{(p)} \right]^{-1} \bar{\mathrm{T}}_{u,u}^{(p - 1)}, \label{eq:redh1}\\
	\bar{\mathrm{R}}_{u,d}^{(p)} &= \tilde{r}_{u,d}^{(p)} + \tilde{t}_{u,u}^{(p)} \bar{\mathrm{R}}_{u,d}^{(p-1)}\left[ \bar{\mathrm{I}} - \tilde{r}_{d,u}^{(p)} \bar{\mathrm{R}}_{u,d}^{(p - 1)} \right]^{-1} \tilde{t}_{d,d}^{(p)},\\
	\bar{\mathrm{R}}_{d,u}^{(p)} &= \bar{\mathrm{R}}_{d,u}^{(p - 1)} + \bar{\mathrm{T}}_{d,d}^{(p - 1)} \tilde{r}_{d,u}^{(p)}\left[ \bar{\mathrm{I}} - \bar{\mathrm{R}}_{u,d}^{(p - 1)} \tilde{r}_{d,u}^{(p)}  \right]^{-1} \bar{\mathrm{T}}_{u,u}^{(p-1)},\\
	\bar{\mathrm{T}}_{d,d}^{(p)} &= \bar{\mathrm{T}}_{d,d}^{(p-1)} \left[ \bar{\mathrm{I}} - \tilde{r}_{d,u}^{(p)} \bar{\mathrm{R}}_{u,d}^{(p - 1)}  \right]^{-1} \tilde{t}_{d,d}^{(p)}, \label{eq:redh4}
\end{align}
where the recursion is initialised by
\begin{equation}
	\bar{\mathrm{S}}^{(-1)} =  \left[\begin{array}{cc}
	\bar{\mathrm{I}} & 0 \\
	0 & \bar{\mathrm{I}}	
\end{array}\right].
\end{equation}
This calculation, carried out at each interface, results in a recursive algorithm for $\bar{\mathrm{S}}^{N}$.

\section{Simulations}
\label{SecSimulations}
\subsection{Planar Reflectance}
\begin{figure*}[t]
	\includegraphics[width=0.8\textwidth]{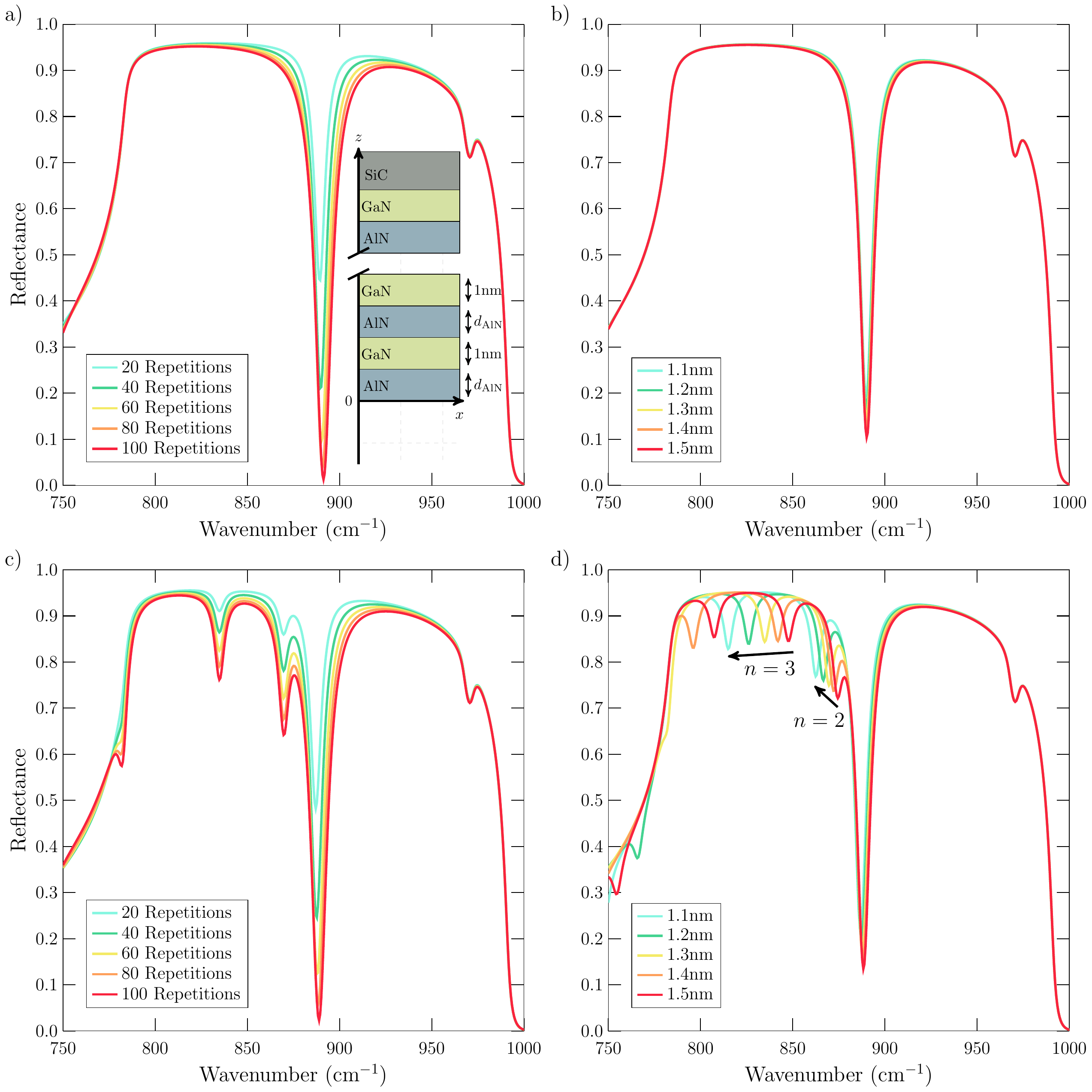}
	\caption{\label{fig:Fig1} The reflectance, calculated at an incident angle of $65^{\circ}$, for an AlN/GaN superlattice on a deep 4H-SiC substrate. a. (c.) Shows the result for GaN layer thickness $1.3$nm for multiple sublattice repetitions in the local (nonlocal) case. b. (d.) Shows the result for $50$ sublattice repetitions over a range of AlN/ GaN layer thicknesses in the local (nonlocal) case.}	
\end{figure*}
The formalism presented can in principle be applied to any spatially nonlocal anisotropic material system, including metallic ones in the limit $\beta_{\mathrm{T}} \to 0$, where the matrices are appropriate truncated to exclude transverse matter modes \cite{Ciraci2013}. In the remainder of this Paper we focus on polar dielectrics in the mid-infrared spectral region, specifically we analyse heterostructures comprised of GaN, AlN and 4H-SiC, a material system with uniaxial anisotropy where crystal hybrids have been previously demonstrated \cite{Ratchford2019} and where a treatment of optical non-locality, albeit using an approximate isotropic theory, has been shown to be necessary to determine the optical response \cite{Gubbin2020b}. To describe an experimental system, grown by molecular beam epitaxy, the $z-$axis of the coordinate system is aligned with the crystal axis. The transverse phonon frequency matrix is therefore of the form
\begin{equation}
	 \bar{\omega}_{\mathrm{T}} = \left(\begin{array}{ccc}
\omega_{\mathrm{T}, \perp} & 0 & 0 \\
0 & \omega_{\mathrm{T}, \perp} & 0 \\
0 & 0 & \omega_{\mathrm{T}, \parallel}	
\end{array}\right),
\end{equation}
and analogously for the other material parameters,
where $\parallel$ denotes the quantity along the crystal axis and $\perp$ those perpendicular to it. All simulations utilise the parameters given in Table~\ref{fig:Tab1}.\\
Note that it was previously demonstrated how, in order to fit experimental data, it is necessary to adjust the nominal thickness of the nonlocal layers \cite{Gubbin2020b}. This reflects the shortcomings of the continuum model in Eq.~\ref{eq:IonEOM}, which does not account for the change in the local environment of ions near the edge of each layer, whose motion is constricted by the presence of the adjacent material. This leads to an effective shortening of each layer length of an amount of the order of the lattice parameter when compared to the nominal layer thickness. Having previously demonstrated the effect can be easily accounted for by adjusting the layer thickness through comparison with experimental data, and in order to avoid introduction of additional parameters, we do not consider such an effect in this Paper. This means that the parameters used will lead to an underestimate of nonlocal shifts, particularly in very thin layers.\\

We begin by studying the reflectance from an AlN/GaN superlattice grown on a deep 4H-SiC substrate under plane-wave illumination. The superlattice, schematised in the inset of  Fig.~\ref{fig:Fig1}a, is characterised by AlN layer thickness $d_{\mathrm{AlN}}$, GaN thickness of $1$nm, and repetition number $R_N$. In Fig.~\ref{fig:Fig1}a (c) we plot the TM polarised reflectance for AlN thickness $d_{\mathrm{AlN}} = 1.3$nm over a range of $R_N$ in the local (nonlocal) case. The incident angle of impinging light is fixed at $65^{\circ}$. The dominant feature in the reflectance is the large hump from $800-970\mathrm{cm}^{-1}$ which is due to the 4H-SiC substrate's Reststrahlen region \cite{Passler2017}, the dip near $970\mathrm{cm}^{-1}$ arises from the 4H-SiC axial LO phonon mode \cite{Paarmann2016}. In the local case there is a strong dip in the reflectance around $890\mathrm{cm}^{-1}$, corresponding to the Berreman mode, near the zone-centre LO phonon frequency in the AlN \cite{Berreman1972, Passler2018}. As the number of repetitions of the AlN/GaN sublattice grows the depth of this dip increases as photons spend more time in the absorbing AlN layers, propagating down to, and returning from the 4H-SiC substrate. In the nonlocal case (Fig.~\ref{fig:Fig1}c) this dip red-shifts slightly because the nonlocal calculation considers the LO phonon supported by the thin AlN film to have finite wavevector. As this wavevector is real the LO phonon modes in the AlN film are quantised, with frequencies
\begin{equation}
	\omega_{n} = \sqrt{\omega_{\mathrm{L}, \parallel}^{(\mathrm{AlN})\, 2} - \left(\frac{n \pi \beta_{\mathrm{L}}^{(\mathrm{AlN})}}{d_{\mathrm{AlN}}}\right)^2},
	\label{eq:discretelo}
\end{equation}
where $\omega_{\mathrm{L}, \parallel}^{(\mathrm{AlN})}$ is the AlN zone-centre phonon frequency along the z-direction and $n$ is an integer describing how many antinodes the quantised phonon mode contains across the layer \cite{Gubbin2020b}. Note that as the wavevector in the in-plane direction, defined by that of the impinging photonic radiation, is small we have assumed the LO mode frequency only depends on out-of-plane phonon velocity $\beta_{\mathrm{L}}$. In the nonlocal case two further dips are observed in the reflectance around $760\mathrm{cm}^{-1}$ and $850\mathrm{cm}^{-1}$. These correspond to the $n = 2$ and $n = 3$ LO modes of the AlN films.\\
It is informative to consider the effect of changing the AlN thickness $d_{\mathrm{AlN}}$ for fixed repetition number. The results using $R_N = 50$ are shown for the local case in Fig.~\ref{fig:Fig1}b for thicknesses in the range $1.1 - 1.5$nm. Increasing the layer thickness results in a very small decrease in the reflectance near the AlN zone-centre LO phonon frequency, due to the increased optical path length through the AlN. In the nonlocal case in Fig.~\ref{fig:Fig1}d the effect is more visible. As predicted by Eq.~\ref{eq:discretelo} an increase in AlN thickness $d_{\mathrm{AlN}}$ results in a strong red-shift of all discrete phonon frequencies away from the zone-centre frequency with decreasing film thickness. Following Eq.~\ref{eq:discretelo} it is clear that higher order modes disperse faster, this is apparent from the motion of the $n=2, \; n=3$ modes in Fig.~\ref{fig:Fig2}d. Angstrom scale changes of the layer thickness result in a visible change in the heterostructure absorption spectrum, demonstrating high tuneability and sensitivity to the best of our knoweledge until now only achieved in similar systems using confined modes \cite{Berte2018,Dubrovkin2020}.\\

\subsection{Surface Phonon Polaritons}
\begin{figure}
	\includegraphics[width=0.5\textwidth]{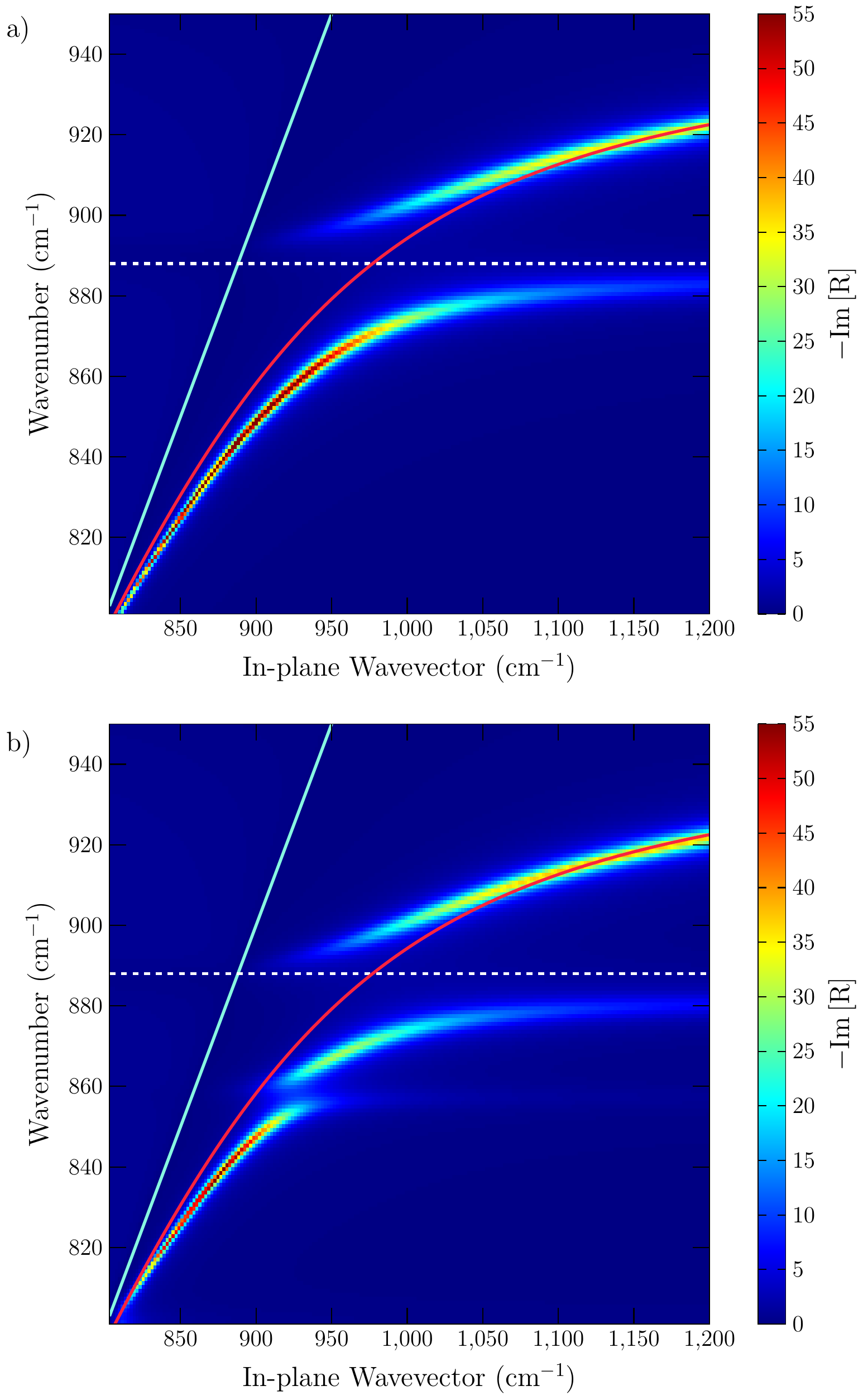}
	\caption{\label{fig:Fig2} Reflectance maps for an AlN/GaN superlattice on a deep 4H-SiC substrate. The lattice is parameterised by layer thickness $d_{\mathrm{AlN}} = 1$nm and contains $50$ repetitions. Results are shown using a. a local dielectric theory, b. the full nonlocal one. In these plots the zone-centre AlN LO phonon frequency is shown by the white dashed horizontal line, the bare 4H-SiC surface phonon polariton dispersion by the red solid line, and the light line is cyan solid.}	
\end{figure}
In addition to calculating the plane wave reflectance of planar polar dielectric heterostructures our model allows for calculation of the reflectance coefficient outside the light-line in the bounding medium. In this region poles in the reflectance correspond to guided modes of the structure \cite{Barnes1998}, which in polar dielectric systems are surface phonon polariton (SPhP). Although these modes are optically inaccessible due to the wavevector mismatch they can be coupled utilising, for example, a prism \cite{Paarmann2016} or by fabricating a grating \cite{Greffet2002}.\\
Compared to plasmonic modes in metals SPhPs have very narrow linewidths \cite{Khurgin2015}. This has made them an interesting platform for the study of strong coupling physics, wherein SPhPs are coupled to different light and matter resonances in order to exploit the unique features of the resulting hybrid excitations. Resonances exploited in these studies include localised phonon polariton modes in nanoresonators \cite{Gubbin2016}, plasmonic nano-rods \cite{Huck2016}, grating modes \cite{Qiang2019}, bulk zone-folded optical phonons \cite{Gubbin2019}, and the epsilon-near-zero (ENZ) mode of a thin polar film \cite{Passler2018}. In the latter case strong coupling occurs around the film's zone-centre LO phonon frequency, at which its dielectric function goes to zero. These modes have been proposed as a promising candidate for mid-infrared field confinement and nonlinear optics due to the strong out-of-plane field enhancement afforded by the Maxwell boundary conditions on the out-of-plane displacement field \cite{Passler2019}. In the following we show that the quantisation of LO phonon modes through Eq.~\ref{eq:discretelo} leads to the emergence of multiple hybrid ENZ modes around the quantised LO phonon frequencies, and that these modes are highly sensitive to the heterostructure layer thicknesses.\\
We calculate the reflectance for the AlN/GaN superlattice on a deep 4H-SiC substrate shown in the inset of Fig.~\ref{fig:Fig1} for $d_{\mathrm{AlN}} = 1$nm and $R_N = 50$. The local result is shown in Fig.~\ref{fig:Fig2}a. In the absence of the superlattice the 4H-SiC substrate supports a surface phonon polariton mode, whose dispersion is shown by the red solid line. The presence of the superlattice causes this to split into two strongly coupled polariton branches, separated by the AlN LO phonon frequency (illustrated by the dashed white line). At large in-plane wavevector the lower polariton is predominantly an ENZ mode, as its frequency approaches the asymptotic LO phonon.\\
In the nonlocal case shown in Fig~\ref{fig:Fig2}b the main polariton branches are almost unshifted. This can be understood from Fig.~\ref{fig:Fig1}d, where we see that for the chosen thickness the film is insufficiently thin for the first LO mode $\omega_1$ (Eq.~\ref{eq:discretelo}) to have shifted appreciably from the zone-center longitudinal mode. Still, the nonlocal nature of the system leads to an additional anti-crossing around $\omega_{2} \approx 865 \mathrm{cm}^{-1}$. The surface phonon polariton is thus strongly coupled to two quantised LO modes in the film. The wavevector dependance of the AlN dielectric function has led to the emergence of a second ENZ mode, demonstrating an additional route to frequency tuneable ENZ physics \cite{Passler2019}.\\
\begin{figure}
	\includegraphics[width=0.5\textwidth]{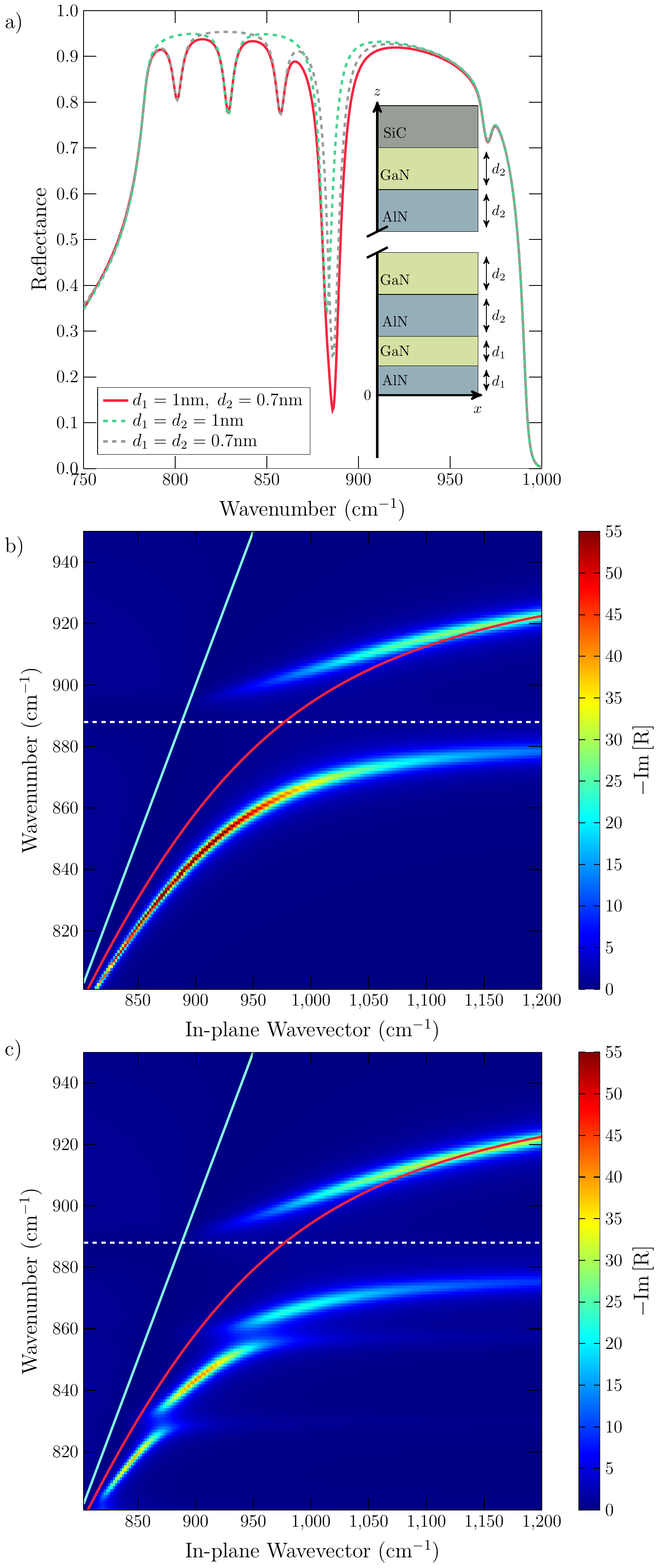}
	\caption{\label{fig:Fig3} a. Reflectance at $65^{\circ}$ for AlN/GaN superlattices defined by $R_N = 50$ for different values of the GaN and AlN layers $d_1$ and $d_2$. Other plots show the imaginary component of the reflectance for the composite structure in the b. local and c. nonlocal cases. In these plots the zone-centre AlN LO phonon frequency is shown by the white dashed horizontal line, the bare 4H-SiC surface phonon polariton dispersion by the solid red line, and the light line is cyan solid.}
\end{figure}
This result means the heterostructure absorption is extremely sensitive to small changes in the resonant layer thickness and opens up interesting possibilities to achieve broadband absorption. We then consider a more complex superlattice, schematised in the inset of Fig. \ref{fig:Fig3}a, whose sublattice is still composed of AlN/GaN but contains two distinct thicknesses $d_1,\;d_2$
\begin{align}
	\mathrm{sublattice} &= \mathrm{AlN} \; d_1 / \mathrm{GaN}\; d_1 /  \mathrm{AlN} \; d_2 / \mathrm{GaN}\; d_2 ,\nonumber
\end{align}
from which the superlattice can be constructed as
\begin{equation}
	\mathrm{superlattice} = \prod_{p = 1}^{R_N} \mathrm{sublattice}. \nonumber
\end{equation}
We calculate the reflectance at $65^{\circ}$ for a superlattice parameterised by $ R_N = 50, d_1 = d_2 = 1$nm and another by $R_N = 50, d_1=d_2= 0.7$nm. Results are shown by dashed lines in Fig.~\ref{fig:Fig3}a. All structures exhibit a dip corresponding to the Berreman mode on the blue side of $\omega \approx \omega_{\mathrm{L}, \parallel}^{\mathrm{AlN}}$ The $1$nm superlattice also exhibits strong reflectance dips at $\omega \approx 855 \mathrm{cm}^{-1}, \omega \approx 800 \mathrm{cm}^{-1}$, these correspond to the $n = 2,\; 3$ quantised LO phonons in the AlN films. The $0.7$nm superlattice exhibits a reflectance dip at $\omega \approx 825 \mathrm{cm}^{-1}$ which corresponds to its $n = 2$ quantised LO mode. Reflectance from the composite heterostructure with $R_N = 50, \; d_1 = 1\mathrm{nm},\;d_2 = 0.7 \mathrm{nm}$ is shown by the red solid line in Fig.~\ref{fig:Fig3}a, it exhibits reflectance dips at all the localised phonon frequencies of each sublattice.\\
Now we calculate the guided modes of the composite heterostructure. Results in the local case are shown in Fig.~\ref{fig:Fig3}b. These are very similar to the results in Fig.~\ref{fig:Fig2}a, with a marginal closing of the anti-crossing. In the local case the only thing that matters is the sum thickness of the AlN/ GaN heterostructure, which determines the coupling between the surface phonon polariton and the Berreman mode of the film. In the nonlocal case Fig.~\ref{fig:Fig3}b the result is more complex. There is a slight red shift in the asymptotic lower polariton frequency, reflecting the shift of $\omega_1$ from $\omega_{\mathrm{L}, \parallel}^{\mathrm{AlN}}$. Two additional anti-crossings are visible around the $n = 2$ frequencies of each sublattice: the heterostructure now supports three hybrid ENZ modes. Such a simulation provides a further demonstration of the extreme sensitivity of the heterostructure optical response to nanometer-sized features.

\section{Conclusions}
\label{SecConclusions}
In this work we have extended theories of optical nonlocality in polar systems to account for the optical response of the bi- and uni-axial media typically utilised in surface phonon polaritonics. Compared to the isotropic case previously considered \cite{Gubbin2020a, Gubbin2020b} this results in an expansion of the mode space in the dielectric, necessitating the consideration of 5 unique modes. We demonstrated that in order to calculate the optical response of a nonlocal multilayer heterostructure it is necessary to follow a scattering matrix approach to avoid numerical instabilities arising from the mismatch between the phonon and photon wavelengths. \\
We applied our model to the study of a variety of GaN/AlN systems, experimentally accessible heterostructures often used in polar nanophotonics \cite{Ratchford2019}, calculating both plane-wave reflectance and guided mode dispersion relations. In doing so we demonstrated how nonlocality can be exploited to dramatically improve the sensitivity of the optical response of the heterostructure,  with small changes to the layer structure resulting in the appearance of new modes due to coupling to quantised LO phonons in the layers.\\
A solid understanding of this tuneability will become increasingly important as surface phonon polaritonics moves further toward the exploitation of nanoresonators for the realization mid-infrared optoelectronic devices. Simulating these systems using ab-initio approaches will be of extreme difficulty, as typical polar nanoresonators are $100$nms across, while the layers of a crystal hybrid are typically of nanometer scale. We are confident our semi-analytical approach, together with the underlying numerical code \cite{Code} will provide a useful tool for the growing community interested in the science and technology of phonon polaritons and mid-infrared nanophotonics.\\

\section{Acknowledgements}
\label{SecAck}
S.D.L. is supported by a Royal Society Research fellowship. The authors acknowledge support from the Royal Society Grant No. RGF\textbackslash EA\textbackslash181001.

\bibliographystyle{naturemag}	
\bibliography{bibliography}

\appendix
\section{The Effective Stress Tensor}
In Eq.~\ref{eq:IonEOM} we introduce the effective stress tensor $\bar{\tau}$, whose exact form is dependent of the symmetry of the lattice. The components of $\bar{\tau}$ can be derived considering the related fourth-rank elasticity tensor $c_{i,j,k,l}$ as in Eq.~\ref{eq:strain} and \ref{eq:elacons}, associated with the crystal free energy
\begin{equation}
	F = \frac{1}{2} c_{i,j,k,l} s_{i,j} s_{k,l}.
\end{equation}
The elasticity tensor $c_{i,j,k,l}$ has $3^4 = 81$ components, however in order to ensure invariance of the Lagrangian, as seen through the free energy, $c_{i,j,k,l}$ must satisfy
\begin{equation}
	c_{i,j,k,l} = c_{j,i,k,l} = c_{k,l,j,i}, \label{eq:sym}
\end{equation}
meaning that only $21$ of these components are unique.\\
The number of components of $c_{i,j,k,l}$ necessary to describe a crystal depends on the lattice symmetry. Here we give results for two common classes of materials used in polar optics, namely cubic and hexagonal crystal structures. A full discussion of the symmetries of the elasticity tensor can be found in Ref.~\cite{Bona2004} and the unique components for a variety of symmetries can be found in Ref.~\cite{ClaytonBook}.\\

\subsubsection{Cubic Symmetry}
In the cubic case the crystal is symmetric under inversion with respect to the three cartesian axis, so transformations like $r_i \to - r_i \; (i = 1, 2, 3)$ must leave the free energy invariant. This means components with unmatched indices like $\lambda_{1,1,1,2}$ vanish. Furthermore all coordinate axis are fourfold symmetry axis, leaving three different non-zero components
\begin{align}
	\lambda_1 = c_{1,1,1,1} &= c_{2,2,2,2} = c_{3,3,3,3}, \\
	\lambda_2 = c_{1,1,2,2} &= c_{1,1,3,3} = c_{2,2,3,3} = \dots,\\
	\lambda_3 = c_{1,2,1,2} &= c_{2,1,1,2} = c_{1,3,1,3} = \dots,
\end{align}
where the dots indicate other permutations of indices through Eq.~\ref{eq:sym}. These components relate to the velocities given in the main body of the paper through
\begin{align}
	\lambda_1 &= \rho \beta_{\mathrm{L}}^2,\\
	\lambda_2 &= \rho \left(\beta_{\mathrm{L}}^2 - 2 \beta_{\mathrm{T}}^2\right),\\
	\lambda_3 &= \frac{\rho}{2} \beta_{\mathrm{C}}^2.
\end{align}
Finally through Eq.~\ref{eq:elacons} the components of the effective stress tensor can be derived
\begin{align}
	\tau_{1,1} &=  \rho \left[\beta_{\mathrm{L}}^2 s_{1,1} + \left(\beta_{\mathrm{L}}^2 - 2 \beta_{\mathrm{T}}^2 \right) \left(s_{2,2} + s_{3,3}\right) \right],\\
	\tau_{2,2} &=  \rho \left[\beta_{\mathrm{L}}^2 s_{2,2} + \left(\beta_{\mathrm{L}}^2 - 2 \beta_{\mathrm{T}}^2 \right) \left(s_{1,1} + s_{3,3}\right) \right],\\
	\tau_{3,3} &=  \rho \left[\beta_{\mathrm{L}}^2 s_{3,3} + \left(\beta_{\mathrm{L}}^2 - 2 \beta_{\mathrm{T}}^2 \right) \left(s_{1,1} + s_{2,2}\right) \right],\\
	\tau_{1,2} &= \tau_{2,1} = \frac{\rho}{2} \beta_{\mathrm{C}}^2 s_{1,2},\\
	\tau_{1,3} &= \tau_{3,1} = \frac{\rho}{2} \beta_{\mathrm{C}}^2 s_{1,3},\\
	\tau_{2,3} &= \tau_{3,2} = \frac{\rho}{2} \beta_{\mathrm{C}}^2 s_{2,3}.
\end{align}

\subsubsection{Hexagonal Symmetry}
The crystals considered in the main text (4H-SiC, AlN and GaN) all have hexagonal symmetry, with point group 6mm. For these materials there are five unique components in the tensor $c_{i,j,k,l}$, given by
\begin{align}
	\lambda_1 &= c_{1,1,1,1} = c_{2,2,2,2},\\
	\lambda_2 &= c_{3,3,3,3}, \\
	\lambda_3 &= c_{1,1,2,2} = c_{2,2,1,1},\\
	\lambda_4 &= c_{1,1,3,3} = c_{2,2,3,3} = \dots,\\
	\lambda_5 &= c_{1,3,1,3} = c_{2,3,2,3} = \dots,\\
\end{align}
where the dots indicate other permutations of indices through Eq.~\ref{eq:sym}. The component $c_{1,2,1,2}$ is also non-zero but relates to the above coefficients through
\begin{equation}
	c_{1,2,1,2} = \frac{1}{2} \left(\lambda_1 - \lambda_3\right).
\end{equation}
Taking the growth axis along the $z$-direction (component 3) we can define
\begin{align}
	\lambda_1 &=  \rho \beta_{\mathrm{L}, \perp}^2,\\
	\lambda_2 &=  \rho \beta_{\mathrm{L}, \parallel}^2,\\
	\lambda_3 &=  \rho \left(\beta_{\mathrm{L}, \perp}^2 - 2 \beta_{\mathrm{T}, \perp}^2\right),\\
	\lambda_4 &=  \rho \left(\beta_{\mathrm{L}, \parallel}^2 - 2 \beta_{\mathrm{T}, \parallel}^2\right),\\
	\lambda_5 &= \frac{\rho}{2} \beta_{\mathrm{C}}^2,
\end{align}
and see that the new parameters correspond to the emergence of different velocities for transverse and longitudinal phonons propagating parallel and perpendicular to the growth axis.\\
The components of the stress tensor are finally given by
\begin{align}
	\tau_{1,1} &=  \rho \biggr[\beta_{\mathrm{L}, \perp}^2 s_{1,1} + \left(\beta_{\mathrm{L},\perp}^2 - 2 \beta_{\mathrm{T},\perp}^2 \right) s_{2,2}  \nonumber \\
	& \quad \quad \quad +\left(\beta_{\mathrm{L}, \parallel}^2 - 2 \beta_{\mathrm{T}, \parallel}^2 \right) s_{3,3} \biggr],\\
	\tau_{2,2} &=  \rho \biggr[\beta_{\mathrm{L}, \perp}^2 s_{2,2} + \left(\beta_{\mathrm{L},\perp}^2 - 2 \beta_{\mathrm{T},\perp}^2 \right) s_{1,1} \nonumber \\
	& \quad \quad \quad +\left(\beta_{\mathrm{L}, \parallel}^2 - 2 \beta_{\mathrm{T}, \parallel}^2 \right) s_{3,3} \biggr],\\
	\tau_{3,3} &=  \rho \left[\beta_{\mathrm{L}, \parallel}^2 s_{3,3} + \left(\beta_{\mathrm{L},\perp}^2 - 2 \beta_{\mathrm{T}, \perp}^2 \right) \left(s_{1,1} + s_{2,2}\right) \right],\\
	\tau_{1,2} &= \tau_{2,1} = 2 \rho \beta_{\mathrm{T}, \perp }^2 s_{1,2},\\
	\tau_{1,3} &= \tau_{3,1} = \frac{\rho}{2} \beta_{\mathrm{C}}^2 s_{1,3},\\
	\tau_{2,3} &= \tau_{3,2} = \frac{\rho}{2} \beta_{\mathrm{C}}^2 s_{2,3}.
\end{align}

\section{The Nonlocal Matrix}
We calculate the nonlocal contribution to the lattice equation of motion Eq.~\ref{eq:fullmatrix} by finding the matrix $\bar{\eta}$ which satisfies
\begin{equation}
	\bar{\eta} \mathbf{X} = \nabla \cdot \bar{\tau}.
\end{equation}
Under the limitations discussed in the main text, namely that the $y$-component of the wavevector is zero and that all fields are homogeneous along $y$, the non-zero components of $\bar{\eta}$ are given by
\begin{align}
	\eta_{1,1} &= -\frac{\omega^2}{c^2} \left(
	\beta_{\mathrm{L}}^2 \zeta^2 + \frac{\beta_{\mathrm{C}}^2}{2}  q^2\right),\\
	\eta_{1,3} &= \eta_{3, 1} = -\frac{\omega^2}{c^2} \left(\beta_{\mathrm{L}}^2 - 2 \beta_{\mathrm{T}}^2 + \frac{\beta_{\mathrm{C}}^2}{2}\right) q \zeta,\\
	\eta_{2,2} &= - \frac{\omega^2}{c^2} \frac{\beta_{\mathrm{C}}^2}{2} \left(\zeta^2 + q^2\right),\\
	\eta_{3,3} &= - \frac{\omega^2}{c^2} \left(\beta_{\mathrm{L}}^2 q^2 + \frac{\beta_{\mathrm{C}}^2}{2}\zeta^2 \right).
\end{align} 

\section{Truncation of the Equations of Motion}
To reduce the full electro-mechanical equations of motion in Eq.~\ref{eq:Full} to a manageable system of equations we eliminate the redundant field components. This proceeds as follows. Initially utilising the 3rd, 5th, and 9th rows of $\bar{\mathrm{M}} \mathbf{F} = 0$ we can write
\begin{align}
	&c \epsilon_0 \mathrm{E}_z + \zeta \mathrm{H}_y + \mathrm{P}_z = 0,\\
	&- q \mathrm{E}_x +  \zeta \mathrm{E}_z + c \mu_0 \mu_{y} \mathrm{H}_y = 0,\\
	&\epsilon_0 \left(\epsilon_{\infty, z} - 1\right) \mathrm{E}_z - \mathrm{P}_z + \alpha_{z} \mathrm{X}_z = 0.
\end{align}
to eliminate $\mathrm{H}_y,\; \mathrm{E}_z, \; \mathrm{P}_z$ as 
\begin{align}
	\mathrm{H}_y &= - c \frac{\epsilon_0 \epsilon_{\infty, z} q \mathrm{E}_x +  \alpha_{z} \zeta \mathrm{X}_z }{\zeta^2 - \epsilon_{\infty, z} \mu_{y} } \label{eq:sub1},\\
	\mathrm{E}_z &= \frac{\zeta q \mathrm{E}_x + \alpha_{z} \mu_{y} \mathrm{X}_z / \epsilon_0}{ \zeta^2 -\epsilon_{\infty, z} \mu_{y} },\label{eq:subEz}\\
	\mathrm{P}_z &=  \frac{\epsilon_0 \left(\epsilon_{\infty, z} - 1 \right) \zeta q \mathrm{E}_x + \alpha_{z} \left(\zeta^2 - \mu_{y}\right) \mathrm{X}_z}{\zeta^2 -  \epsilon_{\infty, z} \mu_{y}}.
\end{align}
Now utilising the 4th and 6th row of $\bar{\mathrm{M}} \mathbf{F} = 0$ the remaining in-plane and out-of-plane magnetic fields can be written as
\begin{align}
	\mathrm{H}_x  &= - \frac{q}{c \mu_0 \mu_{x}} \mathrm{E}_y,\\
	\mathrm{H}_z  &= \frac{\zeta}{c \mu_0 \mu_{y}} \mathrm{E}_y.
\end{align}
Finally utilising rows 1 and 2 we can find the remaining components of the polarisation field as
\begin{align}
	\mathrm{P}_x &= - \frac{\epsilon_0 \left(\zeta^2 + \epsilon_{\infty, z} q^2  - \epsilon_{\infty, z} \mu_{y}\right)\mathrm{E}_x +  \alpha_{z}  q \zeta \mathrm{X}_z }{\zeta^2 - \epsilon_{\infty, z} \mu_{y} },\label{eq:subPx}\\
	\mathrm{P}_y &= \epsilon_0 \left[\frac{q^2}{\mu_{x}} + \frac{\zeta^2}{\mu_{z}} - 1 \right]\mathrm{E}_y. \label{eq:subn}
\end{align}
The remaining equations, rows 7, 8, 10, 11 and 12 of $\bar{\mathrm{M}} \mathbf{F} = 0$ can be written explicitly as
\begin{align}
	\epsilon_0 \left(\epsilon_{\infty, x} - 1 \right) \mathrm{E}_x  - \mathrm{P}_x +  \alpha_{x} \mathrm{X}_x &= 0,\label{eq:red1}\\
	\epsilon_0 \left(\epsilon_{\infty, y} - 1 \right) \mathrm{E}_y  -  \mathrm{P}_y +  \alpha_{y} \mathrm{X}_y &= 0,\label{eq:red2}\\
	\frac{\alpha_{x}}{\rho\omega^2} \mathrm{E}_x + \left[1 +  \frac{i \gamma_x}{\omega} - \frac{\omega_{\mathrm{T,x}}^2}{\omega^2}\right] \mathrm{X}_x\nonumber\\
	 - \frac{1}{\omega^2} \left[\eta_{1,1} \mathrm{X}_x + \eta_{1,2} \mathrm{X}_y+\eta_{1,3} \mathrm{X}_z \right] &= 0 \label{eq:red3},\\
	\frac{\alpha_{y}}{\rho\omega^2} \mathrm{E}_y + \left[1 + \frac{i \gamma_y}{\omega} - \frac{\omega_{\mathrm{T,y}}^2}{\omega^2}\right] \mathrm{X}_y \nonumber\\
	 - \frac{1}{\omega^2} \left[\eta_{2,1} \mathrm{X}_x + \eta_{2,2} \mathrm{X}_y+\eta_{2,3} \mathrm{X}_z \right] &= 0 ,\label{eq:red4} \\
	\frac{\alpha_{33}}{\rho\omega^2} \mathrm{E}_z + \left[1+ \frac{i \gamma_z}{\omega} - \frac{\omega_{\mathrm{T,z}}^2}{\omega^2} \right] \nonumber\\
	 - \frac{1}{\omega^2} \left[\eta_{3,1} \mathrm{X}_x + \eta_{3,2} \mathrm{X}_y+\eta_{3,3} \mathrm{X}_z \right] &= 0 ,\label{eq:red5}
\end{align}
where we divided rows 10, 11 and 12 by $\omega^2$ for consistency. Finally we can substitute the eliminated field components. Using Eq.\ref{eq:subPx} into Eq.~\ref{eq:red1}
\begin{multline}
	\epsilon_0 \left[ \epsilon_{\infty, x}  + \frac{  \epsilon_{\infty, z} q_z^2  }{\zeta^2 - \epsilon_{\infty, z} \mu_{y} }\right] \mathrm{E}_x\\
	+  \alpha_{x} \mathrm{X}_x + \frac{\alpha_{z}  q_z \zeta \mathrm{X}_z }{\zeta^2 - \epsilon_{\infty, z} \mu_{y} } = 0\label{eq:redsub1},
\end{multline}
Eq.\ref{eq:subn} into Eq.~\ref{eq:red2}
\begin{equation}
	 \epsilon_0 \left[\epsilon_{\infty, y} - \frac{q_z^2}{\mu_{x}} - \frac{\zeta^2}{\mu_{z}}  \right]\mathrm{E}_y +  \alpha_{y} \mathrm{X}_y = 0 \label{eq:redsub2},
\end{equation}
and Eq.~\ref{eq:subEz} into Eq.~\ref{eq:red5}
\begin{multline}
	\frac{\alpha_{z}}{\rho\omega^2} \frac{\zeta q_z \mathrm{E}_x}{ \zeta^2 -\epsilon_{\infty, z} \mu_{y} }\\
	 + \left[1 + \frac{i \gamma_z}{\omega} - \frac{\omega_{\mathrm{T,z}}^2}{\omega^2} + \frac{\mu_{y} \epsilon_{\infty, z}}{\omega^2} \frac{ \omega_{\mathrm{L},z}^2 -  \omega_{\mathrm{T},z}^2}{ \zeta^2 -\epsilon_{\infty, z} \mu_{y} }\right] \mathrm{X}_z \nonumber\\
	 - \frac{1}{\omega^2} \left[\eta_{3,1} \mathrm{X}_x + \eta_{3,2} \mathrm{X}_y+\eta_{3,3} \mathrm{X}_z \right]  = 0 \label{eq:redsub3},
\end{multline}
where we utilised the definition of the lattice effective charge density \cite{Gubbin2020a}
\begin{equation}
	\alpha_{z}^{2} = \epsilon_0 \epsilon_{\infty, z}\rho \left(\omega_{\mathrm{L},z}^2 - \omega_{\mathrm{T},z}^2\right).
\end{equation}
Now we have the quadratic equation of motion Eq.~\ref{eq:redeom}. We can easily calculate the non-zero elements of the three matrices entering this equation. For the matrix $\bar{\mathrm{A}}$ we find
\begin{align}
	\mathrm{A}_{1,1} &= \frac{ \epsilon_0  \epsilon_{\infty, z}}{\zeta^2 - \epsilon_{\infty, z} \mu_{y} },\\
	\mathrm{A}_{2,2} &= - \frac{\epsilon_0}{\mu_{x}},\\
	\mathrm{A}_{3,3} &= \frac{\beta_{\mathrm{C}}^2}{2 c^2},\\
	\mathrm{A}_{4,4} &= \frac{\beta_{\mathrm{C}}^2}{2 c^2},\\
	\mathrm{A}_{5,5} &= \frac{\beta_{\mathrm{L}}^2}{c^2}.
\end{align}
For $\bar{\mathrm{B}}$ we find
\begin{align}
	\mathrm{B}_{1,5} &= \frac{\alpha_{z} \zeta}{\zeta^2 - \epsilon_{\infty, z} \mu_{y} },\\
	\mathrm{B}_{3,5} &= \mathrm{B}_{5,3} =  \frac{1}{c^2} \left(\beta_{\mathrm{L}}^2 - 2 \beta_{\mathrm{T}}^2 + \frac{\beta_{\mathrm{C}}^2}{2}\right) \zeta,\\
	\mathrm{B}_{5,1} &= \frac{\alpha_{z} }{\omega^2  \rho}\frac{\zeta }{ \zeta^2 - \epsilon_{\infty, z} \mu_{y} }.
\end{align}
For $\bar{\mathrm{C}}$ we find
\begin{align}
	\mathrm{C}_{1,1} &= \epsilon_0 \epsilon_{\infty, x},\\
	\mathrm{C}_{1,3} &= \alpha_{x} ,\\
	\mathrm{C}_{2,2} &= \epsilon_0\left(\epsilon_{\infty, y} - \frac{\zeta^2}{\mu_{z}}\right),\\
	\mathrm{C}_{2,4} &= \alpha_{y},\\
	\mathrm{C}_{3,1} &= \frac{\alpha_{x}}{\omega^2  \rho},\\
	\mathrm{C}_{3,3} &= 1 + \frac{i \gamma_x}{\omega} - \frac{\omega_{\mathrm{T},x}^2}{\omega^2} + \frac{\beta_{\mathrm{L}}^2 }{c^2} \zeta^2,\\
	\mathrm{C}_{4,2} &= \frac{\alpha_{y} }{\omega^2 \rho},\\
	\mathrm{C}_{4,4} &= 1 + \frac{i \gamma_y}{\omega} - \frac{\omega_{\mathrm{T},y}^2}{\omega^2} + \frac{\beta_{\mathrm{C}}^2 }{2 c^2}\zeta^2,\\
	\mathrm{C}_{5,5} &= 1 + \frac{i \gamma_z}{\omega} - \frac{\omega_{\mathrm{T},z}^2}{\omega^2} + \frac{\beta_{\mathrm{C}}^2 }{2 c^2}\zeta^2 \nonumber \\
	& \quad \quad + \frac{\mu_{y} \epsilon_{\infty, z}}{\omega^2} \frac{ \omega_{\mathrm{L},z}^2 -  \omega_{\mathrm{T},z}^2}{ \zeta^2 -\epsilon_{\infty, z} \mu_{y} }.
\end{align}

After the equation of motion is transformed through Eq.~\ref{eq:lineom} we are left with a new $10 \times 10$ matrix $\bar{\Upsilon}$, whose components are calculable from $\bar{\mathrm{A}}, \; \bar{\mathrm{B}}, \; \bar{\mathrm{C}}$. The non-zero components of $\bar{\Upsilon}$ are given by
\begin{align}
	\Upsilon_{1,6} &= \Upsilon_{2,7} = \Upsilon_{3,8} = \Upsilon_{4,9} = \Upsilon_{5,10} = 1,\\
	\Upsilon_{6,1} &= \epsilon_{\infty, x} \left( \mu_{y}-  \frac{\zeta^2}{\epsilon_{\infty, z}}\right),\\
	\Upsilon_{6,3} &= \frac{\alpha_{x}}{\epsilon_0} \left( \mu_{y}-  \frac{\zeta^2}{\epsilon_{\infty, z}}\right),\\
	\Upsilon_{7,2} &= \mu_{x} \left(\epsilon_{\infty, y} - \frac{\zeta^2}{\mu_{y}}\right),\\
	\Upsilon_{7,4} &= \frac{\alpha_{y} \mu_{x}}{\epsilon_0}, \\
	\Upsilon_{8,1} &= - \frac{2 \alpha_{x} c^2}{\omega^2 \rho \beta_{\mathrm{C}}^2},\\
	\Upsilon_{8,3} &= \frac{2 c^2}{\omega^2 \beta_{\mathrm{C}}^2} \left[\omega_{\mathrm{T}, x}^{2} - \omega \left( \omega + i \gamma_x\right) - \frac{\beta_{\mathrm{L}}^2}{c^2} \zeta^2 \omega^2 \right],\\
	\Upsilon_{9,2} &= - \frac{2 \alpha_{y} c^2 }{\omega^2\rho \beta_{\mathrm{C}}^2},\\
	\Upsilon_{9,4} &= \frac{2 c^2}{\omega^2 \beta_{\mathrm{C}}^2} \left[\omega_{\mathrm{T}, y}^{2} - \omega \left( \omega + i \gamma_y\right) - \frac{\beta_{\mathrm{C}}^2}{2 c^2} \zeta^2 \omega^2 \right],\\
	\Upsilon_{10,5} &= \frac{c^2}{\omega^2 \beta_{\mathrm{L}}^2} \left[\omega_{\mathrm{T}, z}^{2} - \omega \left( \omega + i \gamma_z\right) - \frac{\beta_{\mathrm{C}}^2}{2 c^2} \zeta^2 \omega^2 \right] \nonumber \\
	&\quad - \frac{c^2 \mu_{y} \epsilon_{\infty, z}}{\omega^2 \beta_{\mathrm{L}}^2} \frac{ \omega_{\mathrm{L},z}^2 -  \omega_{\mathrm{T},z}^2}{ \zeta^2 -\epsilon_{\infty, z} \mu_{y} },\\
	\Upsilon_{6,10} &=  - \frac{\zeta \alpha_{z}}{\epsilon_0 \epsilon_{\infty, z}},\\
	\Upsilon_{8,10} &= - \frac{2 \zeta}{\beta_{\mathrm{C}}^2} \left( \frac{\beta_{\mathrm{C}}^2}{2} + \beta_{\mathrm{L}}^2 - 2 \beta_{\mathrm{T}}^2\right),\\
	\Upsilon_{10,6} &=  - \frac{\zeta \alpha_{z}}{\omega^2 \rho \beta_{\mathrm{L}}^2 \left( \zeta^2 - \epsilon_{\infty, z} \mu_{y}\right)},\\
	\Upsilon_{10,8} &= - \frac{\zeta}{\beta_{\mathrm{L}}^2} \left( \frac{\beta_{\mathrm{C}}^2}{2} + \beta_{\mathrm{L}}^2 - 2 \beta_{\mathrm{T}}^2\right).
\end{align}

\section{Sorting of the Eigenmodes}
In order to have a robust algorithm it is necessary to correctly identify each mode in the nonlocal layers of the heterostructure. In lossless media this is a simple task, modes can be sorted by the sign of the real component of their out-of-plane wavevector as
\begin{align}
	\mathrm{Re}\left[q \right] > 0  &\to \text{Down-propagating},\\
	\mathrm{Re}\left[q \right] < 0  &\to \text{Up-propagating}.
\end{align} 
In lossy media $q$ is complex. To avoid unstable growth of solutions we order as
\begin{align}
	\mathrm{Im}\left[q \right] > 0  &\to \text{Down-propagating},\\
	\mathrm{Im}\left[q \right] < 0  &\to \text{Up-propagating}.
\end{align} 
As discussed in the main text in the anisotropic material systems under consideration there are two TO-phonon modes and two photon modes corresponding to TM and TE polarisation states with out-of-plane wavevectors $q_{\mathrm{T}} ^{\mathrm{P}}, q_{\mathrm{ph}}^{\mathrm{P}}$ respectively in which $\mathrm{P}$ labels the polarisation state. There is also a single LO phonon mode with out-of-plane wavevector $q_{\mathrm{L}}$. The LO mode is separated by polarisation. The transverse modes are mixed phonon-photon excitations \cite{Gubbin2020b}. As the phonon dispersion is very slow in reality one will always be matter like and the other light like. We sort by wavevector using $\lvert q_{\mathrm{T}}^{\mathrm{P}}\rvert \gg \lvert q_{\mathrm{ph}}^{\mathrm{P}} \rvert$. Transverse modes are further ordered by polarisation state utilising
\begin{equation}
	\mathrm{C} = \frac{\lvert \mathrm{E}_x \rvert^2}{\lvert \mathrm{E}_x \rvert^2 + \lvert \mathrm{E}_y \rvert^2},
\end{equation}
which is approximately unity for TM modes, when $\mathrm{E}_y = 0$, and zero for TE modes, with $\mathrm{E}_x = 0$. Scalar mode amplitudes in a given layer are written in form $\left[ \mathbf{u}, \mathbf{d}\right]$ where the vector of up-propagating mode amplitudes is $\mathbf{u} = \left[u_{\mathrm{ph}}^{\mathrm{TE}}, u_{\mathrm{ph}}^{\mathrm{TM}},u_{\mathrm{T}}^{\mathrm{TE}}, u_{\mathrm{T}}^{\mathrm{TM}}, u_{\mathrm{L}}\right]$ and that for down-propagating modes is $\mathbf{d} = \left[d_{\mathrm{ph}}^{\mathrm{TE}}, d_{\mathrm{ph}}^{\mathrm{TM}},d_{\mathrm{T}}^{\mathrm{TE}}, d_{\mathrm{T}}^{\mathrm{TM}}, d_{\mathrm{L}}\right]$.

\section{Boundary conditions}
The matrix $\bar{t}^{(p)}$ utilised in Eq.~\ref{eq:tmm1} depends on the fields in layers $p$ and $p+1$. It relates to the single layer interface matrices $\bar{\mathrm{W}}^{(p)}$ through
\begin{equation}
	\bar{t}^{(p)} = \left[\bar{\mathrm{W}}^{(p + 1)} \right]^{-1} \bar{\mathrm{W}}^{(p)}.\label{eq:imat}
\end{equation}
As discussed in the main body of the Paper we apply Maxwell boundary conditions, namely continuity of the electromagnetic fields parallel to the interface $\mathrm{E}_{\parallel}, \mathrm{H}_{\parallel}$, continuity of the full relative ionic displacement $\mathbf{X}$ and of the 3 components of the effective stress
\begin{equation}
	 \bar{\tau} \cdot \hat{\mathbf{z}} = \left(\begin{array}{c}
	\tau_{1, 3}\\
	\tau_{2, 3}\\
	\tau_{3, 3}	
\end{array}\right).
\end{equation}
 The rows of matrix $\bar{\mathrm{W}}^{(p)}$ are these field components for the 5 up- and 5 down-propagating modes in the layer, of the form
\begin{equation}
 \left[	{\mathrm{E}}_x, {\mathrm{E}}_y, {\mathrm{H}}_x, {\mathrm{H}}_y,{\mathrm{X}}_x, {\mathrm{X}}_y, {\mathrm{X}}_z, \tau_{1, 3}, \tau_{2, 3}, \tau_{3, 3}\right]^{\mathrm{T}} ,
\end{equation}
calculated for each mode in the vector $\left[\mathbf{u, d}\right]$ discussed in Appendix D.

\section{Code}
The code is provided in a well documented online repository \cite{Code}. The calculations follow the equations presented in the manuscript. For each unique material in the provided heterostructure the guided mode out-of-plane wavevectors $q$ and mode amplitudes $\left[\mathbf{u, d}\right]$ are calculated through eigenequation Eq.~\ref{eq:eig}. The interface matrix $\bar{\mathrm{W}}_p$, discussed in Appendix E, is then constructed, using the relations in Appendix C to calculate the necessary field components.\\
To calculate the response of a multilayer heterostructure the scattering matrix algorithm presented in Section 2B of the main text is followed. At each interface $z=z_p$ the matrix $\tilde{t}_p$ is constructed through Eq.~\ref{eq:imat}, in conjunction with the eigenvalues $q$ this allows for calculation of the matrix $\tilde{s}_p$ through Eq.~\ref{eq:1sm}. The Redheffer star product recursion Eqs.~\ref{eq:redh1}-\ref{eq:redh4} are then used to find the external scattering coefficients. This process is repeated until the total heterostructure scattering matrix $\bar{\mathrm{S}}_N$ has been calculated, whose block $\bar{\mathrm{R}}_{ud}^{(N)}$ (Eq.~\ref{eq:smm}) yields the heterostructure reflectance.

\end{document}